\documentclass[prl,preprint,superscriptaddress,showpacs,byrevtex]{revtex4}

\usepackage[dvips]{graphicx}
\begin{document}
\title{Dynamics of the chiral phase transition in the  2+1 dimensional Gross-Neveu
model}
\author{Fred Cooper} \email{fcooper@lanl.gov}
\affiliation{T-8, Theoretical Division, Los Alamos National Laboratory,
Los Alamos, NM 87545, USA}
\author{Van M. Savage} \email{vsavage@lanl.gov}
\affiliation{T-8, Theoretical Division, Los Alamos National Laboratory,
Los Alamos, NM 87545, USA}
\affiliation{The Santa Fe Institute, 1399 Hyde Park Road, Santa Fe, NM 87501, USA}

\date{\today}

\begin{abstract}
The phase diagram of the Gross-Neveu (G-N) model in 2+1 dimensions as a function of chemical
potential and temperature has a simple curve separating the broken symmetry and unbroken
symmetry phases, with chiral symmetry being restored both at high temperature and high density.  
We study, in leading order in the $1/N$ expansion,
the dynamics of the chiral phase transition for an expanding plasma
of quarks in the Gross-Neveu model in 2+1 dimensions assuming boost invariant
kinematics. We compare the time evolution of the order parameter (mass of the fermion)
for evolutions starting in the unbroken and broken phases.  The proper time evolution
 of the order parameter
resembles previous results in the 1+1 dimensional G-N model in the same
approximation. The time needed to traverse the transition is insensitive to $\mu$.
\end{abstract}
\maketitle

\label{s1}
The phase structure of QCD at non-zero  temperature and baryon density
is important for the physics of neutron stars and relativistic heavy ion
collisions. The Gross-Neveu model {\cite{ref:GN} has been a quite successful toy model
for understanding the general features of chiral symmetry breakdown. 
In a previous paper \cite{ref:us}  we studied the dynamics of the chiral phase transition in
leading order in $1/N$ in the 1+1 dimensional Gross-Neveu  model. Although the
phase structure in that approximation mimics that of two flavor QCD, 
 in 1+1 dimension the exact theory only has  chiral symmetry breaking at
zero temperature.  In 2+1 dimensions, however, the Gross Neveu model with
$Z_2$ chiral symmetry breaking does indeed have interesting
phase structure in the exact theory as shown by lattice simulations \cite{ref:lattice}. The lattice simulations also show that most of the phase structure can be 
understood from the $1/N$ result. The phase
diagram obtained in leading order in 1/N is also qualitatively correct.  

The Gross-Neveu model in $2+1$ dimensions was studied in the large-N expansion extensively
in Refs.~(\cite{ref:warr}-\cite{ref:klim2}).  The Lagrangian for this model is given by
\begin{equation}
{\cal L}=-i{\bar \Psi_i} \gamma^{\mu}\partial_{\mu}\Psi_i-{1\over 2N}g^2(\bar{\Psi}_i\Psi_i)^2,
\label{eq1.1}
\end{equation}
where the index $i=1 \dots N$. Here the $\gamma^{\mu}$ are the conventional {\em four} dimensional
Dirac gamma matrices, but with $ \mu=0,1,2$. 
By introducing the  auxiliary field 
$
 \sigma=-{ig^2}\bar{\Psi}_i\Psi_i/N~ 
$
one can rewrite this Lagrangian in the alternative form
\begin{equation}
{\cal L}_2=-i\bar{\Psi}_i[\gamma^{\mu}\partial_{\mu}+ \sigma]
\Psi_i
-{N \sigma^2 \over 2g^2}.
\label{eq1.2}
\end{equation}

\begin{equation}
\Gamma =S_{eff}=\int~d^3x{\cal L}_2 [\Psi_{cl},\sigma_{cl}] 
-i N tr\ln S^{-1}[\sigma_{cl}].
\label{eq1.5aa}
\end{equation}
where
\begin{equation}
 S^{-1}[\sigma]= -i[\gamma^{\mu}\partial_{\mu}
+ \sigma]\delta(x-y).
\label{eq1.5a}
\end{equation}

Letting $\Psi_i \equiv \Psi$, $N$ then scales out of the action.  
The effective potential at finite temperature and chemical potential for this model has been discussed in detail
by Rosenstein, Warr, and Park and to leading order in large N the effective potential is expressible as
\begin{eqnarray}
&&V[\sigma, \mu, T] =
{\sigma^2 \over 2 g^2} \nonumber \\
&& - 2\int {d^2k \over (2 \pi)^2} \left[ \omega_k + {1\over\beta} \ln(1+ \exp[-
\beta(\omega_k-\mu)]) + \ln(1+ \exp[-\beta(\omega_k+\mu)]) \right], \nonumber \\ 
\end{eqnarray}
where 
$ \omega_k^2 = |k|^2 + \sigma^2 .$
The derivative can be evaluated explicitly in 2+1 dimensions
\begin{eqnarray}
{\partial V \over \partial \sigma}&&={\sigma \over g^2} - 2 \sigma \int {d^2k \over  \omega_k (2 \pi)^2} 
\nonumber \\
&& +2\sigma \int {d^2k \over  \omega_k (2 \pi)^2} \Biggl(  { \exp [-\beta(\omega_k-\mu)] \over 1+ \exp [-\beta(\omega_k-\mu)]} +{ \exp [-\beta(\omega_k+\mu)] \over 1+
\exp [-\beta(\omega_k+\mu)]} \Biggr)
\end{eqnarray}
The temperature dependent part is finite and is given by 
\[ {\sigma \over  \pi \beta} \ln ( 1+ 2 e^{-\beta | \sigma |} \cosh \mu \beta + e^{- 2 \beta |\sigma|}) \]

The vacuum contribution has a linear divergence, which is related to coupling constant renormalization.
The vacuum structure in the cutoff theory can have two phases depending on whether the bare coupling is 
less than or greater than the critical coupling:
\begin{equation}
g^{-2}_{crit} = 4 \int_{-\Lambda_E}^{\Lambda_E} {d^3p_E \over (2 \pi)^3} { 1 \over p_E^2}
=4\int_{-\Lambda_M}^{\Lambda_M} {d^2p \over (2 \pi)^2} { 1 \over 2|p|} = {2\Lambda_E \over \pi^2} =
{\Lambda_M \over \pi}.
\end{equation}

Since we are only interested  in the phase where there is broken symmetry in the vacuum sector,
we can use the vacuum gap equation to renormalize the
leading order in large-$N$ effective potential. 
The value of $\sigma$ at the minimum of the potential for the vacuum determines the 
Fermion mass 
\begin{equation}
{\partial V \over \partial\sigma}|_{\sigma=m_f}=0= \left( {\sigma\over g^2}+tr
S\right)|_{\sigma=m_f}
\rightarrow m_f=-g^2trS[m_f].
\label{eq1.12aa}
\end{equation}
This leads to the well known gap equation
\begin{equation}
{1 \over g^2} =  4    \int_0^\Lambda ~ {2 p^2 d p \over (2\pi)^2}{1\over p^2+m_f^2}= \frac{\Lambda -
|m_f|}{\pi},
\label{eq:gap}
\end{equation}
where we are using a Euclidean cutoff.  
Using the gap equation, we therefore have that 
\begin{equation}
{\sigma \over g^2} = \sigma \left( \frac{\Lambda -
m_f}{\pi} \right), 
\label{eq1.15}
\end{equation} 
and we  obtain in the vacuum sector
\begin{equation}
{\partial V \over \partial\sigma}=-2\sigma\int~{dp_E ~p_E^2\over \pi^2}\left[{1\over p_E^2+\sigma^2}
-{1\over p_E^2+m_f^2}\right] ={\sigma \over \pi} ( |\sigma|- |m_f| ). 
\label{eq1.18}
\end{equation} 
Integrating we obtain
\begin{equation}
V=  {1\over\pi}\Biggl({|\sigma|^3\over 3}-{m_f \sigma^2\over 2} \Biggr).
\end{equation}

Restoring the effects of finite temperature and chemical potential we have
 
\begin{equation}
V_{eff}={1\over\pi}\Biggl({|\sigma|^3\over 3}-{m_f \sigma^2\over 2} \Biggr)
-{2\over\beta}\int{d^2k\over (2\pi)^2}\Biggl(\ln[1+e^{-{E+\mu\over kT}}]
+\ln[1+e^{-{E-\mu  \over kT}})]\Biggr),
\label{eq1.25a}
\end{equation}
and 
\begin{equation}
{\partial V_{eff}\over \partial\sigma}=
{\sigma \over \pi}\Bigg(|\sigma|-|m_f|+{1 \over \beta} \ln [1+2e^{-\beta |\sigma| } \cosh[\mu\beta]
+e^{-2\beta |\sigma|}]\Bigg).
\label{eq1.25}
\end{equation}

The finite temperature effective mass is the solution of
\begin{equation}
\sigma =m_f - {1 \over \beta} \ln \left[ 1+2e^{-\beta |\sigma| } \cosh[\mu\beta]
+e^{-2\beta |\sigma|} \right]   \label{eq:Tgap}
\end{equation}

If we approach the line of phase transitions from the region
in which chiral symmetry is still intact, we can require
\begin{equation}
{\partial^2 V_{eff}\over \partial\sigma^2}|_{\sigma=0}=0,
\label{eq1.31}
\end{equation} 
which straightforwardly leads to the relation that defines the line of phase transitions,
\begin{equation}
m_f={1\over\beta}\ln [2(1+\cosh[\mu\beta])] \rightarrow \mu = {1 \over \beta}\cosh^{-1} [\frac{e^{m_f
\beta}}{2} -1 ]
\label{eq1.32}
\end{equation}
Using this result the phase diagram of this model is easily constructed and shown in 
Fig.~(\ref{f1}). The phase transition described by $V_{eff}$ being second order. 
The exact phase diagram is found in \cite{ref:lattice} and is quite similar.

\begin{figure}
\centering
\includegraphics{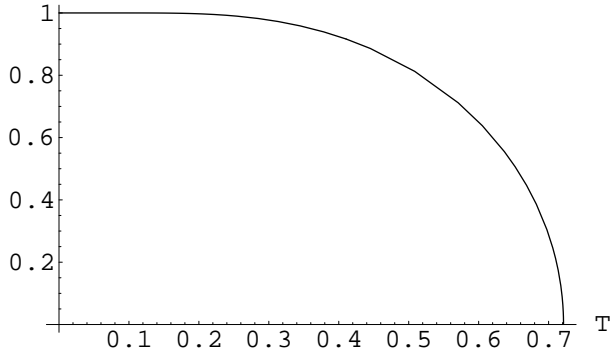}
\caption
{Plot of the line of phase transition given by Eq.(17). The y-axis is the 
chemical potential, and the x-axis is the temperature. In this
plot $m_f=1$ and $k_B=1$.}
\label{f1}
\end{figure}

\section{Dynamics of the phase transition}
\label{s3}

In order to investigate the dynamics of the phase transition in a regime relevant to Relativistic
Heavy Ion Colliders, 
it is convenient to switch to curved space-time coordinates and assume boost invariance for the longitudinal
expansion of an initially highly Lorentz contracted plasma of quarks. This type of assumption has 
been verified recently in the Hartree approximation in scalar field theory (see \cite{ref:luis}). 
Thus we will assume that the expansion possesses hydrodynamical scaling  $x=vt$ \cite{CFS,Bjorken}, where
$x$ is the collisional axis. 

The change of coordinates to a boost invariant frame is given by
$
x=\tau\sinh\eta~{\rm and}~t=\tau\cosh\eta.
$
The corresponding Minkowski line element is
\begin{equation}
ds^2=-d\tau^2+\tau^2 d\eta^2+ dy^2 \to g_{\mu\nu}=(-1,\tau^2,1).
\label{eq2.3}
\end{equation}
Vierbeins are defined by
$
g_{\mu\nu}=V^a_{\mu} V^b_{\nu} \eta_{ab},
$
where $\eta_{ab}=(-1,1,1)$ is the Minkowski metric.
An obvious choice for the vierbein is
\begin{equation}
V^a_{\mu}=(1,\tau,1),
~~~V_a^{\mu}=(1,{1\over\tau}, 1)
 \rightarrow \det V=\sqrt{-g}=\tau.
\label{eq2.6}
\end{equation}

The action for our model in general curvilinear coordinates is
\begin{equation}
S[\Psi,\sigma]=\int d^3x~\det V\bigg({-i\over 2}\bar{\Psi}\tilde\gamma^{\mu} \nabla_{\mu}\Psi
+{i\over 2}(\nabla_{\mu}^{\dagger}\bar{\Psi})\tilde\gamma^{\mu}\Psi-i\sigma\bar{\Psi}\Psi
-{\sigma^2\over 2g^2}\bigg), 
\label{eq2.9}
\end{equation}
where $\tilde\gamma^{\mu}=\gamma^aV_a^{\mu}$, $\nabla_{\mu}=\partial_{\mu}+\Gamma_{\mu}$,
and $\Gamma_{\mu}$ is the spin connection given by \cite{Birrell_Davies,Weinberg}
\begin{equation}
\Gamma_{\mu}={1\over 2}\Sigma^{ab}V_{a\nu}(\partial_{\mu} V_b^{\nu}+\Gamma^{\nu}_{\mu\lambda}V_b^{\lambda})~
{\rm and}~\Sigma^{ab}={1\over 4}[\gamma^a,\gamma^b].
\label{eq2.10}
\end{equation}
The Christoffel symbols have the usual definition,
$\Gamma^{\nu}_{\mu\lambda}=(\partial_\lambda  g^{\nu}_{\mu}+\partial_{\mu}g^{\nu}_{\lambda}
-\partial_\nu  g^{\mu}_{\lambda})/2$.
The only nonzero spin connection is $\Gamma_\eta = - \frac {1}{2} \gamma^0 \gamma^1$. 
~From Eq.~(\ref{eq2.9}), we find that the equation of motion for $\Psi$ is
\begin{equation}
(\tilde\gamma^{\mu}\nabla_{\mu}+\sigma)\Psi=0.
\label{eq2.14}
\end{equation}
Writing this out more explicitly, we obtain
\begin{equation}
(\gamma^2\partial_y+\gamma^0(\partial_{\tau}+{1\over 2\tau})+{\gamma^1\over\tau}\partial_{\eta}
+\sigma)\Psi=0.
\label{eq2.15}
\end{equation}
If we rescale the equation by letting $\Psi=\Phi/\sqrt{\tau}$ and
$\tau={1\over m}e^u$, we obtain,
\begin{equation}
(\tau\gamma^2\partial_y+\gamma^0\partial_{u}+\gamma^1\partial_{\eta}+\tilde\sigma)\Phi
\equiv (\hat\gamma^{\mu}\partial_{\mu}+\tilde\sigma)\Phi=0,
\label{eq2.16a}
\end{equation}
where $\tilde\sigma=\sigma\tau$. 

We expand the Fermion field in terms of Fourier modes
\begin{equation}
\Phi_{\alpha}(u)=\int~{dk_{\eta}~dk_y\over (2\pi)^2}\sum_s \bigg(b_s(k)\Phi_{k,s,\alpha}^+(u)e^{ik\cdot x}
+d_s^{\dagger}(-k)\Phi_{-k,s,\alpha}^-(u)e^{-ik\cdot x}\bigg),
\label{eq2.17}
\end{equation}
where $k\cdot x=k_{\eta}\eta+k_y y$. Note that the dimensions of $b$ and $d$
are in fact $\sqrt{\tau}$, and therefore, it follows that $\Phi^{\pm}$ is dimensionless.
These fields obey the canonical quantization conditions
\begin{equation}
[\Phi_{\alpha}(\tau ,x),\Phi^{\dagger}_{\beta}(\tau ,y)]_+=\delta_{\alpha \beta}\delta^{(2)}(x-y),
\label{eq2.26}
\end{equation}
where $\alpha,\beta$ correspond to the component index and run from 1 to 4. 
We take the standard anti-commutation relations for the creation
and annihilation operators:
\begin{equation}
[b_s(k),b^{\dagger}_{s'}(q)]_+=[d_s(k),d^{\dagger}_{s'}(q)]_+=(2\pi)^2  \delta_{s s'} \delta^{(2)}(k-q).
\label{eq2.27}
\end{equation}
The orthonormality condition for each $k$ is
\begin{equation}
(\Phi^{+}_{k,s}(u))^{\dagger}\Phi^{+}_{k,s'}(u)=
(\Phi^{-}_{-k,s}(u))^{\dagger}\Phi^{-}_{-k,s' }(u)=\delta_{ss'}. 
\label{eq2.31}
\end{equation}

The equations of motion for the mode functions are
\begin{equation}
\left(i\gamma^2 \tau k_y+\gamma^0\partial_{u}+i \gamma^1 k_{\eta}
+\tilde\sigma\right)\Phi^{\pm}_{k,s}(u)=0.
\label{eq2.18}
\end{equation}
This is the equation that we will evolve in the conformal time ($u$) for our simulations.

Explicitly, the equations in component form for the each spin are 
\begin{eqnarray}
(\partial_u-i\tilde\sigma)\Phi_{1,s}+(k_{\eta}-i\tau k_y) \Phi_{2,s}=0\nonumber\\
-(\partial_u+i\tilde\sigma)\Phi_{2,s}+(k_{\eta}+i\tau k_y) \Phi_{1,s}=0\nonumber\\
-(\partial_u+i\tilde\sigma)\Phi_{3,s}-(k_{\eta}-i\tau k_y) \Phi_{4,s}=0\nonumber\\
(\partial_u-i\tilde\sigma)\Phi_{4,s}-(k_{\eta}+i\tau k_y) \Phi_{3,s}=0\nonumber\\
\label{eq:update}
\end{eqnarray}
where $1\to 4$ denote the component. 

For the initial conditions, we take a complete orthonormal
 set of solutions of the Dirac equation
for a fixed initial value of the mass equal to $\sigma(0)$,
\begin{eqnarray}
&&\Phi_{k,s}^{+}(u_0) \equiv y_{k,s}^+ = u_{ks}e^{-i \tilde \omega_k u} \nonumber \\
&& \Phi_{k,s}^{-}(u_0) \equiv y_{k,s}^- = v_{ks}e^{i \tilde \omega_k u}.
\end{eqnarray}
with 
\begin{equation}
u_{ks} = \frac{1}{\sqrt{4 \tilde \omega_k \left(\tilde \omega_k+ \tilde \sigma(0)\right)}}\left(-i \gamma^\mu{\hat k_\mu}+\tilde \sigma(0) \right) \chi_s^+
\end{equation}
\begin{equation}
v_{-ks} = \frac{1}{\sqrt{4 \tilde \omega_k \left(\tilde \omega_k+\tilde \sigma(0)\right)}}\left(i \gamma^\mu{\hat k_\mu}+ \tilde \sigma(0) \right) \chi_s^-
\end{equation}
where 
$\hat k^\mu= \{\tilde \omega_k(0),k_{\eta},\tau k_y \}$, $\tilde \omega_k^2 = {\hat k}^2 + \tilde \sigma^2.$
In what follows we use the explicit representation given in \cite{ref:warr}, 
\begin{equation}
\gamma^{\mu}_{4D}=\gamma^{\mu}_{2D}\left( \matrix{ I  &  0 \cr
0  & -I  \cr}  \right),
\label{eq2.20}
\end{equation}
and 
\begin{equation}
\gamma^{0}_{2D}=i\sigma_3=\left( \matrix{ i  &  0 \cr
0  & -i  \cr}  \right),\gamma^{1}_{2D}=\sigma_1=\left( \matrix{ 0  &  1 \cr
1  & 0  \cr}  \right),\gamma^{2}_{2D}=\sigma_2=\left( \matrix{ 0  &  -i \cr
i  & 0  \cr}  \right). 
\label{eq2.21}
\end{equation}
\begin{equation}
i\gamma^{0}_{4D}=(-1, 1,1,-1).
\label{eq2.22}
\end{equation}
We choose the $\chi_s^{0 \pm}$ as the positive and negative eigenvectors of $i\gamma^{0}_{4D}$. 
\begin{equation}
\chi^{0+}_{1}=\left( \matrix{ 0 \cr 1  \cr 1 \cr 0}  \right)~~
\chi^{0+}_{2}=\left( \matrix{ 0 \cr 1  \cr -1 \cr 0}  \right),~~
\chi^{0-}_{1}=\left( \matrix{ 1 \cr 0  \cr 0 \cr 1}  \right)~~
\chi^{0-}_{2}=\left( \matrix{ 1 \cr 0  \cr 0 \cr -1}  \right).
\label{eq2.24}
\end{equation}
The orthogonality condition for these eigenvectors is
\begin{equation}
(\chi^{0i})^\dag_{s}\chi^{0 j}_{s^\prime}=2\delta_{ij}\delta_{s s^\prime}.
\label{eq2.25}
\end{equation}
where $i= \{+,-\}$.
We also have
\begin{equation}
i\gamma^0\chi^{\pm}_s=\pm\chi^{0\pm}_s,~\gamma^1\chi^{0\pm}_{1,2}=\chi^{0\mp}_{2,1},{\rm and}~
i\gamma^2\chi^{0\pm}_{1,2}=\pm\chi^{0\mp}_{1,2}.
\label{eq2.34}
\end{equation}

\subsection{Effective Mass} 
In lowest order in large $N$, $\sigma$ is the effective mass of the Fermion.  
Its value determines the phase one is in, and it is the order parameter 
for this model in 
this approximation. Let us now obtain a renormalized equation for $\sigma$. 
If we symmetrize $\sigma$

\begin{equation}
\sigma={-i g^2 \over 2}<[\Psi^{\dagger},\gamma^0\Psi]>,
\label{eq2.39}
\end{equation}
 and use $\Psi=\Phi/\sqrt{\tau}$ and Eq.~(\ref{eq2.17}),
we obtain after using the canonical anticommutation relations 
\begin{eqnarray}
&& \sigma=\nonumber\\
&& {g^2 \over 2\tau}\int~{dk_y~dk_{\eta}\over (2\pi)^2} \sum_s \Bigg((1-2N_+(k))
(\Phi^{+}_{k,s}(u))^{\dagger}i\gamma^0\Phi^{+}_{k,s}(u)
+(2N_{-}(k)-1)(\Phi^{-}_{k,s}(u))^{\dagger}i\gamma^0\Phi^{-}_{k,s}(u)\Bigg).\nonumber\\
\label{eq2.44}
\end{eqnarray}
So the equations we need to solve are Eq.~(\ref{eq:update})
for $(\Phi^{\pm}_{k,s}(u))$  and Eq.~(\ref{eq2.44}) for $\sigma$.  To simplify the update we realize that since
\begin{equation}
(\Phi^{+}_{k,s}(u))^{\dagger}i\gamma^0\Phi^{+}_{k,s}(u)= - (\Phi^{-}_{k,s}(u))^{\dagger}i\gamma^0\Phi^{-}_{k,s}(u)
\end{equation}
it is only necessary to solve the Dirac equation for the positive frequency modes.  
As it stands this equation has a {\em linear} divergence.  The most convenient way
of handling the divergence is to use the definition of the unrenormalized coupling constant
(suitably written in terms of mode sums) to relate the bare coupling to the momentum cutoff
and the renormalized mass. That is one uses the lattice version of   
\begin{equation}
{1 \over g^2} =  2    \int_{-\Lambda} ^\Lambda ~ {d^2 p \over (2\pi)^2}{1\over \sqrt{p^2+m_f^2}}.
\end{equation} 
to define the value of $g$ used in Eq.~(\ref{eq2.44}). 

At the initial time $u_0$, Eq.~(\ref{eq2.44}) reduces to the equation for the self-consistent mass at 
the initial temperature and chemical potential. Using the intial time
adiabatic wave functions for $f$, we find for each spin
  
\begin{equation}
(\Phi^{\pm}_{k,s}(u_0))^{\dagger}i\gamma^0\Phi^{\pm}_{k,s}(u_0)=\pm{\sigma\over\omega_k}.
\label{eq2.45}
\end{equation}
Whence,
\begin{equation}
\sigma(u_0) ={2 g^2 \sigma(u_0) \over \tau_0}\int~{dk_y~dk_{\eta}\over (2\pi)^2}\Bigg({1-N_+(k)-N_{-}(k)\over \omega_k}\Bigg),
\label{eq2.46}
\end{equation}
where we should recall that $\omega_k=\sqrt{k\cdot k+\sigma^2}$.

Using the vacuum equation for the gap equation evaluated at $\sigma = m_f$ one obtains the
finite equation for the initial conditions:
\begin{equation}
\sigma(u_o)=m_f+{2 \pi\over\tau_o}
\int~{dk_y~dk_{\eta}\over (2\pi)^2}\Bigg({N_+(k)+N_{-}(k)\over\sqrt{k\cdot k + \sigma(u_o)^2}}\Bigg),
\label{eq2.51}
\end{equation}
where we have used $\tau k_z=k_{\eta}$.
In dimensionless form this is written as
\begin{equation}
\tilde\sigma=\tilde m_f+2\pi
\int~{d^2\tilde k\over (2\pi)^2}\Bigg({N_+(k)+N_{-}(k)\over \tilde\omega_k}\Bigg),
\label{eq2.51aa}
\end{equation}
where $\tilde m_f=\tau m_f$ and $d^2\tilde k = \tau dk_y~dk_{\eta}$.
This is the same equation as the equation we obtained from the finite temperature effective potential Eq.~(\ref{eq:Tgap}). 


The phase transition is quite similar to what we found in 1+1 dimension. We have chosen
three points above the line of phase transitions to study the effect of chemical potential
on the proper time evolution of the transition. We used
  $ \mu =1.0  , T =0.01 $, $ \mu =1.0  , T =0.3 $, and $ \mu =0.8  , T=0.6 $.
In order to break 
the symmetry it is necessary to give $\sigma$ a small initial value. For smaller values than
$\sigma(u_0)=1.e-5$, there was no change in the onset or steepness of the transition. However
for zero or smaller $\sigma(u_0)$ numerical errors could lead to the final value of $\sigma$ being
$\pm 1$. 
We found that for each of these initial conditions, the phase transition occured at almost 
exactly the same value of $u$ and progressed similarly after that.  A comparison of two of 
these simulations using  $ \mu =1.0  , T =0.3 $, and $ \mu =0.8  , T=0.6 $ is shown
in Fig.~(\ref{compare}).  The result for $ \mu =1.0  , T =0.01 $ is quite similar. 
\begin{figure}
\centering
\includegraphics[width=3in]{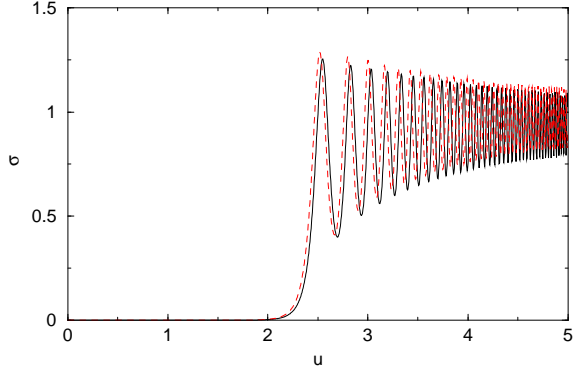}
\caption
{Plot of $\sigma$ vs. $u$ for the second order phase transition.  In this
plot the dashed line is for the initial conditions $T_0 = 0.3$ and $\mu_0 =1.0 $.
The solid line is for $T_0 = 0.6$ and $\mu_0 =0.8 $  We chose an 
initial value $\sigma(u_0)=1.e-5$.}
\label{compare}
\end{figure}

If we instead started in the broken symmetry phase and studied the time evolution we would obtain 
the behavior (for $ \mu= .6  , T = 0.5 $ ) shown in Fig.~(\ref{nophase}). 
We see now that for this case the
proper time needed for the system to relax to the vacuum is shorter, and the transition is not sharp.

\begin{figure}
\centering
\includegraphics[width=3in]{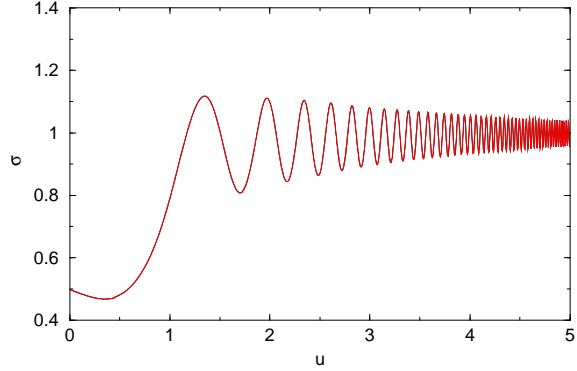}
\caption
{Plot of $\sigma$ vs. $u$ for no phase transition.  In this
plot $T_0 = 0.5$ and $\mu_0 =0.6 $. We chose the self consistent $\sigma(u_0)=.509$.}
\label{nophase}
\end{figure}

\begin{acknowledgments}
FC and VS thank the Santa Fe Institute for their hospitality. 
VS thanks the National Science Foundation for financial support.
We also thank Wenjin Mao for computational assistance.
\end{acknowledgments}

\end{document}